# Language:

# The missing selection pressure

*Jean-Louis DESSALLES*

LTCI, Université Paris-Saclay

dessalles@telecom-paristech.fr - www.dessalles.fr

**Short abstract.** *Why do human individuals compete to provide other individuals with information? Standard ways of explaining the evolutionary emergence of language fail to find a selection pressure for massive information exchange that would concern our species exclusively. I suggest that by talking, human individuals advertise their alertness and their ability to get informed. This behavior evolved as a social signaling device in a context of generalized insecurity that is unique to our species.*

**Abstract.** *Human beings are talkative. What advantage did their ancestors find in communicating so much? Numerous authors consider this advantage to be "obvious" and "enormous". If so, the problem of the evolutionary emergence of language amounts to explaining why none of the other primate species evolved anything even remotely similar to language. What I propose here is to reverse the picture. On closer examination, language resembles a losing strategy. Competing for providing other individuals with information, sometimes striving to be heard, makes apparently no sense within a Darwinian framework. At face value, language as we can observe it should never have existed or should have been counter-selected. In other words, the selection pressure that led to language is still missing. The solution I propose consists in regarding language as a social signaling device that developed in a context of generalized insecurity that is unique to our species. By talking, individuals advertise their alertness and their ability to get informed. This hypothesis is shown to be compatible with many characteristics of language that otherwise are left unexplained.*

Human language bears little resemblance to other primate communication, if only by the magnitude of its characteristics (vocabulary size, volume of exchanges, amount of time devoted to it, intricacy of structure, variety of messages). Accounting for its existence in *Homo sapiens* has been presented as one of the most important (Bickerton 2009) and difficult (Premack 1985; Christiansen & Kirby 2003b) scientific questions. Why is the emergence of language such a conundrum? If, as many authors claim, language is a good thing to have, there must have been a selection pressure for acquiring it. Any general advantage (*e.g.* insulation when temperatures go down) acts as an evolutionary attractor. If such a long range selection pressure existed for language, why were other primate species immune to it? How can we otherwise explain that the selection pressure for language was local enough to concern a few hominin species exclusively?

The purpose of the present paper is first to pose the problem in explicit terms. Many discussions about the evolutionary emergence of language rely on implicit assumptions concerning evolutionary mechanisms that contradict each other. The first section will illustrate this problem by reviewing the various reasons that are invoked when regarding language as "obviously" advantageous. The second section will put the issue in the perspective of the opposition between macro- and micro-evolution. Then, the various 'reasons' why language evolved will be confronted with the reality of language behavior. This will leave us with a new conundrum: it seems that there should have been a selection pressure *against* language as we know it! Or, in other terms, the selection pressure that led to language is still missing. At that point, I will suggest that language evolved in the presence of a form of insecurity that is unique to our species. Being informed became a social value, and language developed as a tool to advertise it.

## 1. Is language "obviously" a good thing to have?

Some authors consider that the selective advantages of language are "enormous" (Chomsky 1975, pp. 40; 2002, p. 148; Hurford 1991a, p. 172; 1991b, p. 293; Penn, Holyoak & Povinelli 2008, p. 123; Brinck & Gärdenfors 2003, p. 495; Vyshedskiy 2014, p. 315) or "considerable" (Bradshaw 1997, p. 100; Savage-Rumbaugh & Lewin 1994, p. 249). For instance:

*"So if one organism just happens to gain a language capacity, it might have reproductive advantages, **enormous** ones." (Chomsky 2002, p. 148)*

*"Language gives humans an **enormous** advantage concerning co-operation in comparison to other species. We view this advantage as a strong evolutionary force behind the emergence of symbolic communication." (Brinck & Gärdenfors 2003, p. 495)*

*"Even in primitive form, such a system of communication would have had **considerable** survival advantages" (Savage-Rumbaugh & Lewin 1994, p. 249.)*

The advantages brought by language are moreover regarded as "obvious" by many authors (Bickerton 1990, p. 156; Berwick & Chomsky 2015, p. 87; Blackmore 1999, p. 99; Blythe & Scott-Phillips 2014, p. 393; Chomsky 1982, p. 46; Donald 1999, p. 148; Ghazanfar & Takahashi 2014, p. 544; Jerison 1973, p. 405; Lieberman 1992, p. 23; Nowak & Komarova 2001; Pinker 1994, p. 367; Pinker & Bloom 1990, p. 712; Wilkins & Wakefield 1995, p. 162). For instance:

*"[…] it is just extraordinarily unlikely that a biological capacity that is highly useful and very valuable for the perpetuation of the species and so on, a capacity that has **obvious** selectional value, should be latent and not used." (Chomsky 1982, p. 46)*

*"The immediate, practical benefits that hominids would have gained from communicating with one another in even the simplest form of protolanguage are **obvious** enough." (Bickerton 1990, p. 156)*

*Vocal language represents the continuation of the evolutionary trend towards freeing the hands for carrying and tool use that started with upright bipedal hominid locomotion. The contribution to biological fitness is **obvious**. (Lieberman 1992, p. 23)*

*"The adaptive significance of human language is **obvious**. It pays to talk." (Nowak & Komarova 2001)*

*There is a fantastic payoff in trading hard-won knowledge with kin and friends, and language is **obviously** a major means of doing so (Pinker 1994, p. 367)*

*It is **obvious**, then, that language is a good thing to have, both for us as individuals and for our species as a whole. (Ritt 2004, pp. 1-2)*

Other authors qualify the advantages of language as "clear" or "uncontroversial" (Allott 1992, pp. 106-107; Christiansen, Dale & Ellefson 2002; Christiansen & Ellefson 2002, p. 338):

*It seems **clear** that humans with superior language abilities are likely to have a selective advantage over other humans (and other organisms) with lesser communicative powers. This is an **uncontroversial** point [...] (Christiansen & Ellefson 2002, p. 338)*

These judgments about the virtues of language are in apparent contradiction with the difficulty of explaining why human beings evolved it. As Burling puts it, we may wonder how "we get from an ordinary primate that could not talk to the strange human primate that can't shut up" (Burling 2005, p. 4). Burling acknowledges the fact that the advantages brought by language may seem obvious to linguists or primatologists, but nevertheless the selective pressures that fostered verbal complexity should not be taken for granted (pp. 182-183). If language is an evolutionary marvel, we must still explain why it did not evolve in other primate species (Hurford 1999; Számadó & Szathmáry 2006). As a recent book title asks, "Why only us?" (Berwick & Chomsky 2015). One possibility is that a selection pressure for developing linguistic skills existed in the biological or ecological context of humans (or hominins) but was absent in the context of other ape species. The next section will explore this possibility.

## 2. Language as a local evolutionary attractor

### 2.1. Natural or cultural selection

Most (but not all) scholars concerned with the origin of human language regard natural selection as the main reason that brought human language to existence. The above quotations (see also section 4) mention a variety of reasons why language would have been *selected*. Some other authors, however, do not consider that natural selection may provide any relevant account of the reasons why language faculties emerged in the first place. Crucial aspects of the language faculty would have emerged in their definitive form just by chance, instead of having evolved gradually through the continued action of natural selection (Bolhuis *et al.* 2014; Chomsky 1975; Tattersall 1998). However, these authors still suppose that the universality and the persistence of language in our species result from its selective advantages (Chomsky 2002, p. 148), which may include improved thinking and improved reasoning in addition to better communication (Berwick & Chomsky 2015, pp. 80, 84).

Natural selection is rightfully irrelevant to authors who do not hypothesize the existence of a specialized biological faculty of language. The idea that language behavior would be a cultural habit like writing or playing chess began to be widely accepted in the nineteenth century as language ceased to be regarded as a godsend (Formigari 1993, p. 150). Its modern proponents consider that crucial aspects of language, such as the ability to process syntax and the willingness to talk, owe their existence to culture. This economy of hypotheses concerning biological endowments specific to language can be found in comparative psychology (Tomasello 1999a, pp. 44, 208; 1999b, p. 526; 2003, p. 109), in anthropology (Knight 2000; Noble & Davidson 1996, p. 214; Schoenemann 2005; Tattersall 2014), in linguistics (Deutscher 2005, p. 19; Dor 2015, p. 190; Van Valin & LaPolla 1997, p. 649), in biology (Jablonka, Ginsburg & Dor 2012), in neuroscience (Arbib 2005, p. 107 and up to a point Deacon 1997, p. 339), in philosophy (Sterelny 2006) and in computer modeling (Christiansen & Chater 2008; Kirby 2000; Steels 2000, p. 2). From the perspective of most of these authors, language (or languages) did emerge through a selective process, though not through a biological one. Cultural selective forces range from vertical transmission and easiness to learn (Christiansen & Chater 2008; Kirby 2000) to cultural selection (Jablonka, Ginsburg & Dor 2012) and communicative efficiency (Christiansen & Chater 2008, p. 543). The comparison with the evolution of writing offers a description of what cultural selection may achieve:

> *We suggest that just as literacy has done during historical time, early language evolution involved socially learned and constructed alterations, adjustments and improvements in communication signs and structures, which came together through historical–cultural evolution. (Jablonka, Ginsburg & Dor 2012, p. 2153)*

Most of the observations that will be made in this section rely on the sole fact that some selective mechanism has been responsible for the emergence of language as we know it, regardless of the biological or cultural nature of the selection. The only alternative to selection is random drift. Random changes in the absence of selective forces are unlikely to produce anything structured, so we will exclude the absence of any selective mechanism from our discussion of language emergence. We will mainly consider the case of biological evolution, but many of our observations should apply, *mutatis mutandis*, to the case of cultural evolution.

## 2.2. Evolutionary peak shift

The first observation we can make about selection is that it produces local optima (Hansen 1997). If language competence, as will be assumed here, results from a selective process, then it must be optimal in some ways. A superficial understanding of this statement may lead to the kind of criticisms that have been (sometimes rightfully) addressed to the so-called adaptationist program. Two fallacies should be avoided (Gould & Lewontin 1979). The first one is to believe that every natural feature must be optimal. By definition, selective mechanisms do not act on neutral features. Moreover, most solutions achieved by selection are a trade-off between conflicting constraints (*e.g.* generation simplicity *vs.* parsing tractability, or learnability *vs.* expressivity), so features may be found to be imperfect for one criterion in isolation. The second fallacy consists in believing that optima produced by selection should be global. Optima are always local. This means that no improvement can be obtained through small changes, independently from any judgment of sub-optimality one may pass from a global and *ex-post* perspective.

Since selection produces local optima, let's suppose that the human species with its ability to master language is located on such an optimum[1]. Figure 1 illustrates the situation. It shows two local optima. The one on the right (point H) represents the situation of language in the human species. The *y*-axis may represent any distinctive feature that would distinguish human language from other forms of animal communication. Here, I represented the amount of information shared among individuals. The chosen dimension is supposed to be correlated with (biological or cultural) fitness. Note that the figure is just a sketch: scales should not be taken too literally. The purpose of this drawing is just to make a logical point concerning the transition from non-language to language.

---

[1] Note that *the species is never what natural selection optimizes*. Natural selection optimizes relative reproductive success within the species. A set of features is locally (Pareto-) optimal only as far as no slight change of these features in some individuals can increase their reproductive success.

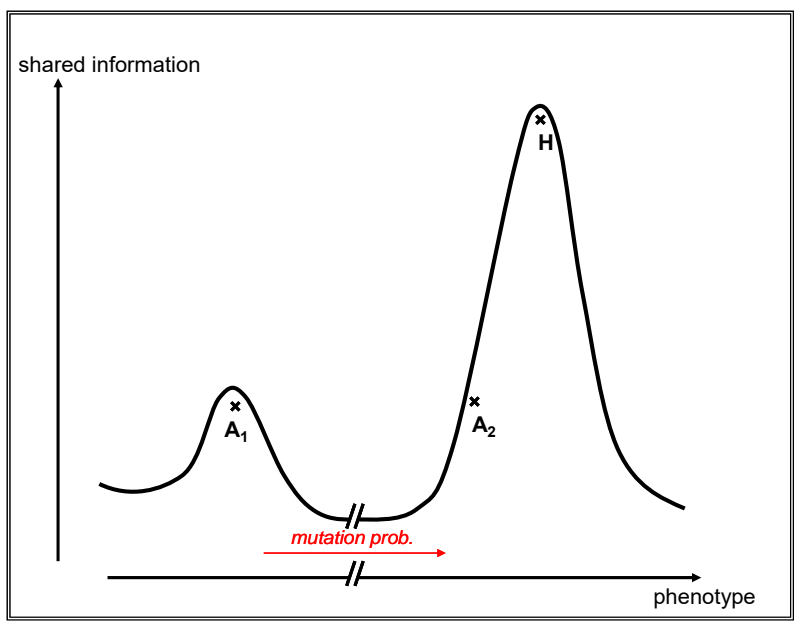

*Figure 1. Schematic representation of an adaptive landscape related to language.*

In this simplified view, non-language is represented by some typical ape communication system (primate species may differ substantially in their communicative behavior, but we may ignore these differences here for our discussion about human language). The question is then whether ape communication (let's say, chimpanzee communication) is better represented by point A1 or by point A2 in Figure 1. If we remember that selection generates local optima, A1 should be the solution. However, A2 is probably more akin to popular views concerning the position of human language in the world of animal communication. If, as suggested by the above quotations, language is such a good thing to have, one may be tempted to believe that other ape species would have evolved it if only given enough time. This thesis was put forward by Sue Savage-Rumbaugh at the Paris Evolang Conference in 2000 and is illustrated by the title of her book, *Kanzi: the ape at the brink of the human mind* (Savage-Rumbaugh & Lewin 1994). The only way of supporting A2 against A1 consists in invoking the slowness of evolutionary processes. According to this view, evolution admittedly tends to produce local optima, but due too its incredibly low speed, no wonder that most species would just be trying hard to find their way to these optima. The slowness of evolution is sometimes used as an argument in the context of language emergence (de Duve 1995, p. 403; Tomasello 1999a, p. 204; Worden 1998, p. 150). These judgments concerning the slow speed of evolution are generally unsupported and might result from the confusion between micro- and macro-evolution.

Micro-evolution means 'evolution under selection pressure'. It is a *hill-climbing* process that corresponds to the transition A2 → H in Figure 1. It is governed by Darwinian laws (or some other selective force for processes such as cultural selection). Micro-evolution *is a rapid process*. Observations (Thompson 1998) and simulations (Dessalles 1996) show that biological change under selection pressure can happen in a few dozens of generations. Even if we allow for hundreds of generations, significant change can occur in a few millennia for a species like ours. Such duration can be considered instantaneous in comparison with the time scale of hominin phylogeny. Computer scientists take advantage of the speed of micro-evolution in techniques like genetic algorithms (Goldberg 1989; Dessalles 1996). The high speed of micro-evolution is due to a phenomenon that John Holland named 'implicit (or intrinsic) parallelism' (Holland 1975). Various mutations are 'tried' in different individuals or lineages in parallel; the most successful ones replicate; thanks to genetic recombination (crossover), they have some probability of ending up together in a same genome; individuals endowed with this genome inherits valuable traits that have been selected in several of their ancestors independently. This phenomenon of implicit parallelism explains why evolution under selection pressure is significantly faster than if mutations had to accumulate one after the other in a single lineage.

Macro-evolution, on the other hand, is a slow process. The macro level is characterized by the phenomenon of *punctuated equilibria* (Eldredge & Gould 1972). This phenomenon is the unavoidable consequence of the action of a rapid selective process. In an idealized view, species rapidly evolve and end up sitting on a local optimum. They may remain there in equilibrium an indefinite amount of time. Then, for whatever reason such as some improbable mutation, some individuals discover a neighboring local optimum. A burst of micro-evolution follows, which corresponds to a punctuation in Eldredge and Gould's schema. This process results in a *peak shift*, such as the transition A1 → H in Figure 1.

This peak shift phenomenon is not the only possibility at the macro-scale, as the 'adaptive landscape' may change through time, due to environmental or ecological perturbations (Estes & Arnold 2007). In this case, the species is expected to follow the local adaptive optimum as it moves. We may call this process *moving peak tracking*.

According to this simplified description of evolutionary change, major modifications would result either from hill-climbing (evolution under selection pressure), adaptive peak shift or moving peak tracking. Things may be of course more complex, and when averaged over very long periods, adaptive changes merge into a combination of the three processes (Estes & Arnold 2007). However, if we focus on one single qualitative change such as the transition to language, we must decide which of hill-climbing, peak shift and peak-tracking offers the best account.

Due to the high speed of micro-evolution, we can rule out the hill-climbing scenario. A primate species located at a point like A2 in the adaptive landscape (Figure 1) would have evolved a communication system resembling language (point H) in a few hundreds or thousands of generations. This means that non-human primate species are in equilibrium (as illustrated by point A1) and are unlikely to be evolving toward more language-like communication. The transition to language therefore appears as a genuine macro-evolutionary event.

The peak-tracking scenario could in theory produce an evolutionary path to language: slowly changing conditions would have brought hominin species to evolve communication systems looking more and more like language. Though this possibility cannot be ruled out, it remains quite unlikely. It would presuppose that language is no more than an amplified version of ape communication, and that at each step during its evolution, the amplification factor would have been fine-tuned to the 'needs' of the time. These alleged changing conditions play the role of a non-parsimonious external cause that led evolution on a definite track. In the absence of any proposal for such environmental conditions that would have led to language by repeated adaptive amplifications of primate-like vocalizations, the peak-tracking scenario remains unsupported.

The peak-shift scenario is the last remaining option: *Language can only be a local attractor*. This conclusion is at odds with traditional descriptions that invoke some long-lasting evolutionary trend toward larger brains (Schoenemann 2006), greater intelligence and eventually language (Lieberman 1992, see above quotation). In principle, 'evolutionary trends' do not occur at the macro-evolutionary level and are only possible during the ephemeral micro-evolutionary bursts. Macro-evolution is apparently unpredictable and non-directed; it has no inertia and no memory. As Gould (1996) convincingly demonstrated, a species has no means to remember where it was coming from as it reached its current equilibrium. This conclusion conflicts with anthropocentric views of evolution that would regard language as an absolute advantage towards which all other mammal species are slowly striving to evolve.

## *2.3. Preadaptations*

Repeated peak shifts may have produced what can still be regarded with hindsight as some sort of evolutionary trend toward language. The idea consists in imagining that hominin species, while hopping from one adaptive peak to another, as do all species, fortuitously came close to a situation where language became evolvable. Reading this 'success story' in retrospect would give an illusion of evolutionary trend. The intermediary steps are often called 'preadaptations' (Christiansen & Kirby 2003a; Hurford 2003a; Wildgen 2004). Numerous candidates have been proposed, including some brain rewiring (Wilkins & Wakefield 1995), the ability to control the vocal tract (Savage-Rumbaugh & Lewin 1994), the ability to combine meaningless sounds (Sereno 2005), the ability to use symbols and to combine them (Deacon 1997), the mastery of conceptual complexity (Schoenemann 2005), the ability to process recursive structures (Berwick & Chomsky 2015; Chomsky 1975), the ability to imitate (Arbib 2005; Donald 1999; Zlatev 2014), the ability to share a focus of attention (Tomasello 2003; 2006), the ability to read others' beliefs and desires (Baron-Cohen 1999), the mastery of complex social relationships (Cheney & Seyfarth 2005; Worden 1998), an increase of social group sizes (Dunbar 1996), the existence of widespread cooperation (Nettle 2006) or a major social organization shift (Knight 2008).

The very notion of preadaptation comes with two requirements: that the preadaptation was hard to evolve, and then that language was easier to evolve once the preadaptation was installed. This hard-before-easier-after scenario is however questionable on both sides. Let's consider the 'hard-before' side. The name 'preadaptation' presupposes that the corresponding quality evolved independently from language, rather than being a consequence of it. But why did the preadaptation evolve only in our lineage? Merely asserting its "obvious" advantages deepens the mystery instead of solving it.

*Mimesis would have provided **obvious** benefits, allowing hominids to expand their territory, extend their potential sources of food, and respond more effectively as a group to dangers and threats. (Donald 1999, p. 148)*

After having invoked a preadaptation (such as Donald's mimesis) that is unique to hominins, one needs two independent evolutionary accounts instead of one. The situation illustrated in Figure 1 has to be repeated twice, first to explain why the transition to the preadaptation occurred only once, and then to explain how language could evolve from the preadaptation. In particular, one must show how the preadaptation is locally optimal and why non-human primates did not evolve it. The difficulty augments of course if one imagines that several of these preadaptations were necessary for language to eventually emerge. Postulating preadaptations should therefore be done with parsimony and as the last resort.

Preadaptations make sense only if they made the emergence of language more likely. They are generally claimed to have "cleared the way" for its evolution or, alternatively, their absence would have "prevented" other species from evolving language:

[About increased orofacial motor control and enhanced social intelligence] *These two innovations, conspicuously absent in nonhuman primates, may have been crucial to clear the way for the emergence of language in modern humans. (Slocombe & Zuberbühler 2005, p. 1783)*

*The conclusion is thus that it is the lack of bodily mimesis […] that prevents non-human creatures from evolving both cumulative culture and language. (Zlatev 2014)*

In most cases, preadaptations are presented as conducive to language, as if a breach in the firewall surrounding the language adaptive peak had been opened. The diversity of candidates should alert us about the lack of support for such hypotheses: each would-be preadaptation must be shown to be the decisive one, at the exclusion of all other candidates. In this respect also, proposing a preadaptation for language appears difficult and should not be done lightly.

To summarize, the transition to language is necessarily a macro-evolutionary event, due to the speed of micro-evolution. As a consequence, it is impossible or, at best, non-parsimonious, to speak of an 'evolutionary trend' toward language. Language must be a *local* evolutionary attractor, and nothing predestined our lineage to evolve it. Figure 1 might be somewhat misleading in this respect. The transition from A1 to A2 and then to H may appear inevitable in the long run, as there is only one great adaptive peak in A1's vicinity. This conclusion, of course, corresponds to no reality and would just result from an *ex post* bias. One should rather imagine that the A1 peak is surrounded by many other adaptive peaks at various distances and various heights in a high-dimensional space. The fact that only one dimension and one neighboring peak are featured in Figure 1 is just for the sake of simplicity.

This observation that language is not the outcome of some long-ranging evolutionary trend comes with several corollaries. Various features of primate communication seem to bear non-trivial resemblance to human language. For instance, two meaningful signals may be combined to mean something new (Arnold & Zuberbühler 2006; Ouattara, Lemasson & Zuberbühler 2009). Due to the inexistence of long-term evolutionary trends, such features, whatever their interest as instances of convergent evolution, should not be analyzed as precursors or embryonic forms of the corresponding language features. No species can be regarded as an imperfect 'draft' in anticipation or in the direction of the human species. This holds for apes, but also for hominin species such as *Homo erectus* or *Homo ergaster*. Nothing predestined them to become *sapiens*.

Another consequence of the locality of macro-evolutionary attractors is that contrary to what is sometimes asserted (Schoenemann 2005), we should expect a modular architecture of the language faculty. Each past species was in equilibrium, which means that each of its selected features was locally optimal. If, as suggested by Derek Bickerton (1990), *Homo erectus* used protolanguage to communicate, then protolanguage must have been locally optimal for its function. Protolanguage was not a draft of language and did not announce it. The succession of equilibria lets us expect qualitative differences between successive species, each step being fully functional. An analogy is offered by open-source computer programs that come with various additional and successive *modules* or add-ons. One may choose to add new functionalities and to get rid of old features. Similarly, new behavioral characteristics emerging in a biological species after an adaptive peak shift are expected to correspond most of the time to qualitative, not quantitative, changes. Darwin famously claimed that "the difference in mind between man and the higher animals, great as it is, certainly is one of degree and not of kind." (1871, ch. IV). Though Darwin was certainly right to think that all intermediary forms did exist at some point in time[2], there is no reason to conclude, as he apparently did, that every difference must be quantitative. In the case of language, the fact that changes might have been gradual (Christiansen & Chater 2008, p. 503; Corballis 2014; Gibson 1994; King 1996) does not entail that the outcome is only quantitative variation. Nature offer numerous examples of qualitative innovations produced through gradual change. A peak shift like the A1 → H transition (Figure 1) predicts qualitative change: some previous characteristics that were positively selected in A1 may have lost their adaptive value, while new characteristics that were neutral or detrimental become advantageous at point H. If language evolved through several adaptive peak shifts, as suggested by the protolanguage hypothesis, then we should expect several qualitative differences between human language and other ape communication systems. We consider some of these differences in the next section.

## 3. The strangeness of human language

Some of the most remarkable facts about language are relative to the way it is *used*. These facts are rarely considered in evolutionary accounts, despite their relevance to the issue of language origin. One reason is that linguists generally choose to split the language faculty into a linguistic component (anything up to context-independent meaning) and a communicative component, and quite often choose to ignore the latter. I consider the dichotomy artificial, as it may obfuscate the functional dependencies between the different modules that make up the language faculty. Moreover, choosing to ignore language use is like studying the heart while obstinately refusing to consider the way it pumps blood. This attitude is regrettable, as some facts about language use might well turn out to be the key to understanding why language emerged in our species. Table 1 lists some of these facts. As we will see, many of them turn out to be hard to explain.

---

[2] Some authors suggested (Chomsky 1975; Bickerton 1990) that the transition to language was due to some single improbable mutation that would have generated qualitative differences in one single step. Even if correct (despite being improbable), this kind of hypothesis is irrelevant to our discussion which deals with the adaptiveness of language. The question of why the mutation has been successful would remain.

Table 1. *Facts to be explained about human communicative behavior.*

| | |
|---|---|
| SPADV | For language to exist, speakers must get some advantage from speaking. |
| COST | Language is a costly apparatus and a costly behavior. |
| INCONS | Many topics addressed in spontaneous conversation are about inconsequential matters. |
| OFFER | Human beings are talkative. They speak more than ten thousand words per day on average, often in a competitive way, striving to be heard. |
| VOCAB | Individuals acquire plethoric vocabularies. Adults understand tens of thousands of different words. |
| NOAUDEX | People willingly talk to various audiences, even to people they don't know, and most often to several people simultaneously, with little audience exclusion. |
| GEN | Language is a generalized behavior: very few people refrain from talking and remain systematically silent. |
| GDRN | Language is gender neutral. Women and men show similar conversational behavior. |
| ABNORM | Events reported in spontaneous conversations are most often about abnormal situations, *i.e.* events or states of affairs that are presented as unexpected by speakers. |
| POINTING | Young children spontaneously point to unexpected situations, even before they acquire language. |
| SYNTAX | All human individuals are able to master languages that use central recursion. |
| PRED | Meaning is in part expressed in predicative form. |

**Speakers' advantage.** Many accounts of language existence highlight the benefit that listeners may get from acquiring information, while ignoring that language can only exist if speakers have some incentive to speak.

**Cost.** The cost of language may be overlooked if one focuses on marginal costs exclusively. The energy required to utter one single sentence is indeed negligible. This should not hide the fact that language means a huge investment for human beings (Miller 2000, p. 360). Costs include the amount of time devoted to it: language activity takes up from one fifth to one third of one's waking time (Dunbar 1998; Mehl and Pennebaker 2003). These impressive figures reveal that far from being a marginal habit, language is central to any human life, from nearly birth to death. Language also requires considerable time and sometimes risks to acquire or generate relevant information (Miller 2000, p. 360; Reio Jr. *et al.* 2006). Humans must also bear the cost of having a low larynx (Lieberman 1992) and the cost of supporting a large brain (Aiello 1997) to memorize hundreds of thousands of past experiences worth telling (Dessalles 2007a) in addition to the content of past conversations (Norrick 2000).

**Inconsequentiality.** What do people do with words? Classical philosophical essays (Austin 1962; Searle 1969) suggested that language is primarily used to perform actions. Observation of spontaneous language offers a different picture. Some authors studied spontaneous language (*i.e.* casual conversation occurring 'in the wild', under uncontrolled conditions) and paid attention to the content of conversations, to what people talk about. Their work reveals several surprising facts which are at odds with the idea that language would be an action-oriented tool. First, many topics are about futile matters that are inconsequential to participants (Dunbar, Duncan & Marriott 1997). For instance, they may talk about a soccer team playing in the third division having reached the semi-finals of the national Cup (my own corpus, 12.04.2000) or they may wonder which animals do eat hot chili pepper in the wild (my own corpus, 26.09.2000). We will see that the inconsequentiality of human chat is one of the facts that are particularly hard to explain. Another fact revealed by the observation of spontaneous conversation is that people mostly talk about past situations, especially during narratives (Dessalles 2017; Mahr & Csibra, *to appear*; Norrick 2000; Tannen 1984). During story rounds (Tannen 1984, p. 100) people recount past events one after the other, each story triggering the next one. Narratives represent 40% of topics in a corpus of family conversations I analyzed (Dessalles 2017). This amount matches other estimates (Eggins & Slade 1997). This form of past-oriented declarative speech departs from theories in which language would appear as an action-oriented behavior.

**Competitive offer.** Language is a disproportionate behavior. Individuals speak about 16,000 words on average per day, while some may reach 45,000 (Mehl *et al.* 2007). Narratives, arguments and opinions are spontaneously and massively offered during conversation, most of the time without any prompting. This readiness to offer information for free is also observed in technical forums and social networks (Ariely 2008, p. 89; Kwak *et al.* 2010). It is also characteristic of the scientific publication system. The readiness to talk has been compared to a competition in which individuals strive to be heard (Dessalles 1998; Miller 2000, p. 350).

**Vocabulary.** Vocabularies are plethoric. Average individuals understand tens of thousands of different words (often from two of more languages) that they learn at an impressive rate during childhood (Goulden, Nation & Read, 1990).

**No audience exclusion.** Verbal interactions are far from being systematically dyadic; whenever possible, individuals speak to three or more (Dunbar *et al.* 1995), with little or no audience exclusion (Miller 2000, p. 350).

**Generalized behavior.** Observations across various cultures reveal that virtually all individuals engage in talking (Dunbar 1998). The distribution of the number of words spoken per person per day reveals that the vast majority of people do talk (Mehl *et al.* 2007).

**Gender neutrality.** Language is in first approximation gender neutral. Both sexes engage in conversational activities, with no quantitative difference (Mehl *et al.* 2007; Redhead & Dunbar 2013) and at best only slight qualitative differences that may depend on local conditions (Aries & Johnson 1983; Tannen 1994).

**Abnormal situations.** The observation of spontaneous conversation reveals that people systematically talk about *abnormal* situations.

*"We would intuitively reject such introductions as 'Let me tell you something ordinary that happened yesterday…' A narrative that is in fact judged to be ordinary may be rejected after it is told by expressions equivalent to 'So what!'" (Labov & Fanshel 1977, p. 105)*

Situations worth talking about have to be "problematic" (Ochs *et al.* 1992), "different from ordinary experience" (Labov & Fanshel 1977, p. 105), "unexpected, deviant, extra-ordinary, or unpredictable" (van Dijk 1993), "abnormal" (Schank 1979), "odd or unexpected", "rare", "impossible or unheard of", be "the violation of a norm" (Polanyi 1979), "be a low probability event", "depart from expectations" (Agar 2005; see also Davies 1971). Even when exchanging about weather conditions, interlocutors feel obliged to frame trivial observations as abnormal events (unusually long rainy period, erroneous forecast, high temperature for the season). Unexpectedness or abnormality requires that the situation depart from expectations. For instance, one would not expect a third-division team to reach the semi-finals of the national Cup; or one would not expect any animal to enjoy eating hot chili pepper. The notion of unexpectedness has been formalized[3] (Dessalles 2013) and makes correct predictions about acceptability and interest in conversation (Dessalles 2017).

**Declarative pointing.** Bickerton (2009) challenges theories of language emergence to explain the usefulness of the very first dozen of words. But even before that, we should be able to explain the existence of pointing behavior. All human beings, including children by the age of twelve months, draw attention to unexpected events using for instance declarative pointing (Carpenter, Nagell & Tomasello 1998, p. 58). Apes may point, but only in an imperative way to get food for instance (Pika & Liebal 2006; Tomasello 2006). Apes are curious, but they don't apparently share their curiosity, or at least not systematically, contrary to humans.

**Syntax.** One of the most celebrated differences is the human ability to process central embedding (recursion). In the preceding sentence, the noun phrase 'central embedding' is included in the verb phrase 'process central embedding' which is itself embedded in the noun (or determiner) phrase 'the human ability to process central embedding'. The resulting structure is best represented as a binary tree that may grow from any of its branches. This tree may be the apparent outcome of a dynamic process based on operations such as *merge* (Chomsky 1995). Every healthy human individual is able to process central recursion in language, while there is little evidence that non-human species can.

**Predicates.** Language conveys meaning. Let's consider the sentence:

> *The sister of the colleague you met yesterday is a chess champion.*

The meaning of this sentence has several dimensions, including a perceptual one (the addressee may have formed an image of the colleague, and perhaps could 'see' the woman playing chess). It is also tempting to offer a formal representation of the sentence's meaning, as in this Prolog[4] translation:

---

[3] Unexpected situations require complex circumstances to occur. For instance, a third-division team can only reach the semi-finals of the national Cup thanks to a complex conjunction of favorable circumstances. Unexpected situations also require a small amount of information to be distinguished. For instance, the same event would appear less unexpected, and thus less interesting, if it occurred in a foreign country that needs to be specified. These two factors can be characterized using the notion of 'minimum description length'. See www.simplicitytheory.science for definitions, examples and references.

[4] Prolog is a computer language based on predicates and variables. Variables start with a capital letter. All occurrences of the same variable within a clause refer to a same value (which may be unknown).

*meet*(*you, C, yesterday*), *colleague*(*C, me*), *sister*(*S, C*), *champion*(*S, chess*).

This translation makes use of predicates: *meet*, *colleague*, *sister* and *champion*. The nature of these predicates has been widely discussed in linguistics, in philosophy and in computer science. A commonly shared opinion is that predicates predate interpretation, and even pre-exist in fixed form in the mind, as elements of a language of thought. The *ad hoc* choice of predicates in the preceding example and their proximity to linguistic expression should make us suspicious about this idealized perspective[5]. I submitted the idea that predicates are not permanent representations and may be formed on the fly at the time of interpretation (Dessalles 2015). The important fact for our present discussion is that the sentence of our example translates quite naturally as a bag of predicates. This kind of representation can be directly used by a Prolog interpreter to draw inferences and perform reasoning.

The ability to form predicates can be regarded as a fundamental cognitive ability of our species, even if some other species are sometimes granted with an embryonic version of it (Hurford 2003b). Predicates (or the ability to generate them) fulfill two functions that are linked to the communicative role of language. One of them is to receive an attitude such as (dis)belief, desire, undesirability, unexpectedness. The attitude is generally carried by the predicate that lies on top of the syntactic hierarchy. In our example, it would be *champion*(*lucy, chess*). The other function of predicates is to support determination. If you do not know *lucy*, I can replace her name by a variable (*S* in the example) and use another predicate, *sister*(*S, C*), to help you determine who I am talking about. And then I can use yet another predicate, *meet*, to help in the determination of the newly introduced variable *C*. This process is a recursive one, occurring at the semantic level. Syntactic recursion can be seen as a tool for expressing predicates that are recursively used for determination[6] (Dessalles 2007b[2000], pp. 213-6).

Even if we restrict the *explanandum* to the short list of Table 1, determining which selection pressure could lead to a communication behavior with these characteristics remains complicated, as we will see.

## 4. Is language evolutionarily unstable?

According to Terrence Deacon (1997, p. 377), "Looking for the adaptive benefits of language is like picking only one dessert in your favorite bakery". Deacon then mentions a dozen from "the myriad of advantages" that better communication offers, including organizing hunts and planning warfare. The purpose of the present section is not only to question the reality of all these alleged advantages, but even to offer a reversed picture in which the reason why language was selected in the first place will appear as a genuine mystery.

---

[5] One clear description of a language of thought based on pre-existing predicates has been offered by Jerry Fodor (1975). Surprisingly, Fodor himself fought hard to reject the possibility that the language of thought could be anything like an ontology based on definitions or on relations (Fodor 1998; see also Ghadakpour 2003).

[6] The underlying mechanism relies on semantic linking (Dessalles 2007b[2000], p. 216): if two phrases are syntactically connected, then the corresponding predicates must share a variable. With this mechanism, and contrary to Prolog, variables such as *C* or *S* in the example can remain unspoken.

The functions that are supposed to make language beneficial, as "obvious" as they might be (see section 1), vary from author to author. They can be organized in several broad categories that are listed in Table 2. The table shows the various incompatibilities between the selective scenarios and some of the facts mentioned in the previous section. A '+' sign means that the scenario can be claimed to correctly predict the feature, whereas a '–' means that it wrongly predicts its opposite. '0' means that the scenario is compatible with the feature but does not predict it.

The point of the present discussion is not to dismiss these scenarios as outright wrong, but rather to draw attention to the fact that they cannot be claimed to be "obvious". Most of them would require significant amendment to only be considered as potential accounts for the existence of language. Let us comment on this table's lines.

Table 2. *Compatibility of language characteristics with selective scenarios*

| *Characteristic* / Model | SpAdv | Cost | Incons | Offer | Vocab | NoAudEx | Gen | GdrN | Abnorm | Pointing |
|---|---|---|---|---|---|---|---|---|---|---|
| Collective benefit | – | – | – | – | 0 | + | + | + | 0 | 0 |
| Coordination | + | 0 | – | – | – | – | + | 0 | – | – |
| Reciprocity | + | – | – | – | 0 | – | + | + | 0 | – |
| Kin selection | + | + | – | – | – | – | + | 0 | 0 | 0 |
| Manipulation | + | + | – | 0 | 0 | – | + | + | – | – |
| Improved thinking | – | – | – | – | + | + | + | + | 0 | – |
| Social networking | + | + | + | 0 | – | + | + | + | – | 0 |
| Sexual signaling | + | + | + | + | + | – | + | – | – | – |
| Social signaling | + | + | + | + | + | + | – | + | 0 | + |

### *4.1. Collective benefit*

Many accounts of the existence of language invoke some form of collective benefit, either at the group level or at the species level: it would be good to share useful information to increase everyone's knowledge about the material world (Allott 1992, p. 107; Baumeister & Vohs 2002, p. 675; Corballis 2014, p. 51; Corballis and Suddendorf 2007; Deacon 1997, p. 377; Györi 1997; Pinker and Bloom 1990; Ritt 2004, p. 1; Santibáñez 2015; Sterelny 2012, p. 76; Szathmáry & Számadó 2008 p. 40-41). To be eligible as an evolutionary explanation for language, a supposed function must bring an adaptive advantage. More than that, one must show that language corresponds to an evolutionarily stable strategy (ESS). This means that non-speaking (or less-speaking) individuals cannot do better than average in a speaking population. Few of the authors who regard language as "obviously" advantageous (see section 1) did consider this constraint. This is especially true when we consider the many scenarios based on collective benefit in relation to the speaker's advantage constraint (SPADV). These scenarios wrongly predict that speakers would benefit from remaining silent and from merely taking advantage of others' information. This constitutes a special case of the so called 'Tragedy of the commons'. Scenarios that value information pooling consider that information has material value. Giving away useful information for free is therefore an altruistic act. An information pooling population would be invaded by free-riders who benefit from accessing pooled information without contributing to it. Scenarios merely invoking collective benefit do not, as they stand, explain how language behavior can be an ESS. Most other scenarios have been imagined by authors precisely to overcome this difficulty.

Some theoretical models such as group selection (Wilson & Sober 1994) or gene-culture interactions (Choi & Bowles 2007; Richerson & Boyd 2005) show that collective benefit may have an evolutionary impact. These models produce relatively weak selection pressures and require strong discrepancies between groups. They can hardly been invoked to explain the emergence of language, not only because human groups or cultures do not differ regarding language use (Dunbar 1998), but also because language is costly (Table 1). The COST constraint is a problem for any account based on collective benefit: costs endured by individuals should be compared, not with a share of the collective benefit, but with the variation of that share induced by their participation (Sumpter & Brännström 2008). Participating in a collective effort is profitable only in the presence of a strong leveraging effect, when the *marginal* return of this participation outweighs its cost. There is virtually no such leveraging effect in the case of language. Individuals who would unilaterally stop contributing to collective knowledge would still benefit from the same information pool.

The information pooling scenario faces a further difficulty with the INCONS constraint. The scenario relies on the fact that information has some material value. The fact that human conversation is replete with inconsequential topics is at odds with the idea that shared information is so valuable and would benefit group members to such an extent that it would compensate for the cost of language. Even more mysterious is the OFFER constraint. "People compete to say things. They strive to be heard." (Miller 2000, p. 350) This makes no sense within any of the 'information collectivism' scenario.

### *4.2. Coordination*

In coordination situations, participants switch from a disorganized equilibrium to another Nash equilibrium that benefits each of them (Skyrms 2004; Sumpter & Brännström 2008). Language would help in this transition (Bickerton 2009; Brinck 2004; Carruthers 1996, p. 231; Gärdenfors 2004; 2006; Nowak and Komarova 2001; Sterelny 2006; Snowdon 2001; Tallerman 2014, p. 322). Does language fit in with this role of facilitator? If language were geared to organizing action, human conversation topics would be mainly about immediate or future plans. This is not what we observe. Most conversation topics are not instrumental (Redhead & Dunbar 2013) and deal with past situations (Dessalles 2017; Dunbar, Duncan & Marriott 1997; Mahr & Csibra, *to appear*; Norrick 2000). The scenario also conflicts with the INCONS, OFFER, VOCAB and NOAUDEX constraints: conversations should deal with serious matters, words would be few and show little variety (Burling 1986; Redhead & Dunbar 2013), and they should be directed to action partners exclusively. Lastly, the fact that most conversation topics are about abnormal situations (ABNORM, POINTING) makes little sense within the coordination scenario.

### 4.3. Reciprocity

Language has often been described as a form of cooperation (Grice 1975) and conversely cooperation is often invoked to account for the advent of language. In a loose sense, cooperation may refer to one of the previous scenarios, collective sharing and coordination. Otherwise, cooperation means reciprocity. Reciprocal interactions are dyadic situations in which both partners would benefit from cooperating but would get a better short-term payoff by unilaterally defecting. Considerable research efforts have been devoted to finding conditions in which reciprocal cooperation can be stable (Hammerstein 2003; Nowak 2006). Are these conditions relevant to language?

As soon as one considers that language conveys useful information, one may wonder why speakers would willingly provide such information for free. One tentative solution is to consider language exchanges as forms of iterated reciprocity (Fitch 2004; Hurford 2007; Pinker 2003). This idea conflicts with several facts about language (Table 2). Reciprocity is only possible when the benefit of cooperation exceeds by far the cost of cooperating. The high costs associated to language (COST), the fact that many verbal interactions are apparently futile (INCONS), the fact that individuals compete to speak rather than to listen (OFFER) without significant audience discrimination (NOAUDEX) make little sense if language is some sort of information barter. According to the reciprocal schema, overtalkative individuals should be highly appreciated while sporadically speaking ones should be excluded from future conversations. Talking about inconsequential topics should be sanctioned and knowledgeable people (including academics) would be the target of endless requests from the crowd to deliver their wisdom. Most contexts (including academics) offer the converse picture: people strive to be heard or read. On the listeners' side, information may often have a negative price: one is ready to support a cost not to receive it (think of advertisement or spams). And when feedback is provided, it is most often to criticize rather than to thank the speaker (Dessalles 1998; 2017). Reciprocal cooperation may not be what language is about after all (Scott-Phillips 2007).

This does not preclude the possibility that language may be a useful tool to handle the problem of free-riders in reciprocal cooperation by contributing to the emergence of reputations, and that it may have evolved for that reason (Dunbar 1998; 1999; Nowak & Sigmund 2005; Smith 2010). People do indeed comment on cooperative and uncooperative actions, which is part of their gossiping activity (Dunbar 1996; Dunbar, Duncan & Marriott 1997). Reputations of honesty and reliability would be at stake during daily chatter (Scott-Phillips 2011). Language acts, from this perspective, would be meta-cooperative, as they would serve policing purposes. But meta-cooperation faces the same problems as cooperation itself: why bother helping others by warning them about third-parties' uncooperative actions (Power 1998), why so willingly, why with so many words, and why so often by commenting on inconsequential acts?

### 4.4. Kin selection

Altruistic facts represent a puzzle for Darwinian theory. Organisms should be concerned with their own success rather than enduring cost to increase others' success. Unless those others are relatives. One way to explain why individuals feel the urge to provide information to other individuals is to say that language owes its existence to kin selection, either because it was primarily used to teach offspring (Arbib 2005, p. 114; Castro et al. 2004; Fitch 2004) or because its main purpose is to provide hard-won knowledge to close relatives (Lieberman 1992, p. 23; Pinker 1994, p. 367). One problem with these ideas is that hunter-gatherers live in bands in which genetic relatedness is low, too low for kin selection to operate (Hill *et al.* 2011). Another problem is that kin selection can hardly been invoked to explain why language exists at all today. Conveying information that is useful to offspring or relatives requires neither to be talkative (OFFER in Table 2) nor to use tens of thousands of distinct words (VOCAB) (Burling 1986; Dunbar 2003, p. 220). People would not teach futile matters (INCONS) and communication would not be predominantly among non-kin (NOAUDEX). Imagining the possibility that language began to evolve through kin selection in the first place is of little help to understand why it still exists now.

### 4.5. Manipulation

Language can be used to manipulate others' mind and behavior. Unlike animals, we can affect the social and physical world without moving from our chair, merely by using words and by bringing other people to do or believe what we wish them to do or to believe. Did language evolve for this reason (Crespi 2008; Sperber & Origgi 2005)? As suggested in Table 2, this proposal is not fully convincing. If the purpose of talking is to influence others' actions, we should ration our words to speak exclusively to people who can do something tangible for us. This does not fit with the inconsequentiality of our daily conversations (INCONS) and the fact that we are ready to talk to many people whose actions do not affect us (NOAUDEX). The fact that conversations deal primarily with abnormal states of affairs (ABNORM) makes little sense if the aim is to influence others' beliefs and actions on our behalf. Declarative pointing in young children (POINTING) can hardly be analyzed as some kind of manipulation either (in contrast to imperative pointing).

### 4.6. Improved thinking

It has been suggested that the novel cognitive abilities underlying language were initially selected, not for their role in communication, but because they dramatically improved reasoning abilities (Chomsky 2002, pp. 76, 148; Reboul 2007). This assumption leaves several aspects of language unexplained (Table 2). If language originated as an internal cognitive process, why did language become public at all (SPADV), why devote so many resources to this public output (COST), why should individuals talk so much and in a competitive way (OFFER) about inconsequential subjects (INCONS) that are systematically abnormal (ABNORM)? Why do even young children show this communicative behavior (POINTING)? The problem with this internal-first scenario is that it puts us in a situation of explaining two miracles instead of one. It seems more parsimonious to regard the human particular reasoning abilities as a consequence of language rather than as a preadaptation for language. In particular, 'deliberative' reasoning bears all the hallmarks of a by-product of our argumentative competence rather than the other way around (Dessalles 2007b[2000], p. 306; Mercier & Sperber 2011).

### *4.7.  Social networking*

Language is almost by definition a social behavior. One talks with other individuals, preferentially friends. Human beings form coalitions as many social primate species do, but in our case, this is achieved mainly through language (Burling 1986; Dunbar 1996). Did language evolve as a tool to establish and maintain social bonds? According to Dunbar, language replaced grooming in that role. This hypothesis as it stands still falls short of a convincing account. It does not explain the many thousands words of our vocabularies (VOCAB in Table 2), let alone the syntactic and semantic complexities of human languages. At the extreme, it seems that synchronized grunts could do the job of binding coalitions together. The social networking scenario certainly needs additional hypotheses to be compelling[7].

### *4.8.  Sexual signaling*

Geoffrey Miller (2000) championed the idea that language emerged through sexual selection. Our male ancestors would have used language to advertise their intellectual abilities. Our female ancestors would have adopted language as well, but to be able to understand males' performances and also to create enduring couple relationships (Miller's 'Scheherazade' hypothesis). In this scenario, language is more a display device than a way of exchanging tangible information. This explains why topics need not be consequential (INCONS in Table 2). However, the scenario is prone to many criticisms (Fitch 2004). Language should not appear before puberty during development (POINTING) and linguistic abilities of males and females should differ quantitatively and qualitatively (GDRN). Men would talk only to women and conversely (NOAUDEX). And the systematic mention of abnormal situations in spontaneous conversations (ABNORM) is left unexplained.

### *4.9.  Social signaling*

*Social signals* are any conspicuous features or behaviors that are displayed to attract coalition partners. The idea that apparently altruistic acts are in fact signals that serve the performer's interests was introduced by Amotz Zahavi (1975; Zahavi & Zahavi 1997). Sexual signaling is one instantiation of this idea. Another is social signaling. Within Zahavi's framework, these signals are evolutionarily stable if they are reliable indicators of a social quality, *i.e.* a quality that individuals want to find in friends. For instance, mobbing behavior in birds can be interpreted as a way for individuals to advertise their readiness to take risks for the defense of the group (Zahavi & Zahavi 1997; Maklakov 2002). Some mechanisms of 'social selection' (Nesse 2009) are close to the social signaling schema. Can we interpret language as a way of displaying some social qualities to attract social partners? Showing that language is a social display device would be a step toward solving the paradox of its apparent altruistic aspect. It would even explain the competitive offer of information (OFFER in Table 2). Various proposals have been made in this direction. For instance, speakers would gain social status by being eloquent or by talking relevantly (Burling

---

[7] Dunbar proposed such a hypothesis by highlighting the importance of gossip. Burling proposed a social signaling role for language. These hypotheses are discussed in the subsections on reciprocity and on signaling.

1986; Dessalles 1998; Henrich & Gil-White 2000). The social signaling scenario faces two main difficulties. One is to explain the audience's motivation to pay attention to signals emitted by speakers and grant them status for that reason (Henrich and Gil-White suggest that the audience would be eager to learn from authoritative information sources, but this conflicts with the INCONS and OFFER constraints). Another difficulty that has been rarely mentioned comes from the winner-take-all effect (figure 2): best performers get the most part of the social benefit and there is no incentive for the crowd to participate in the signaling game (Dessalles 2014). As a result, only a minority is expected to spend time and resources in talking (GEN in Table 2).

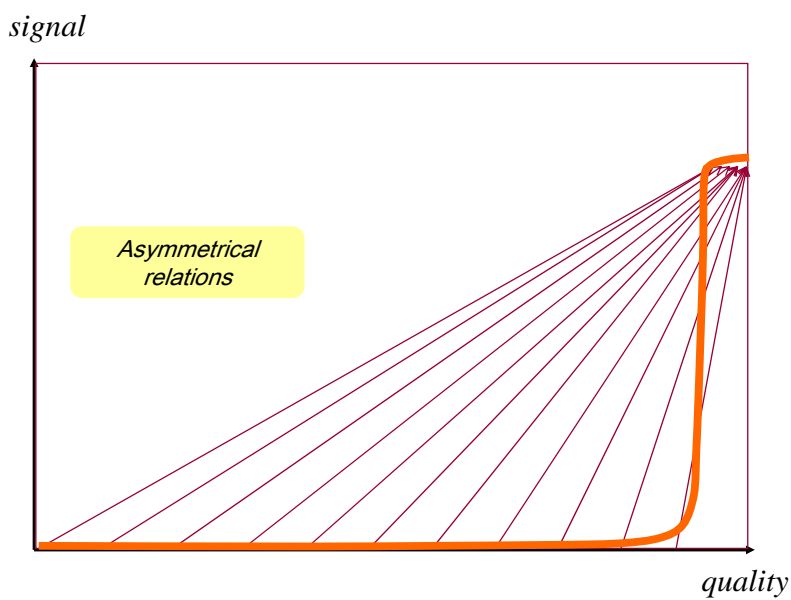

*Figure 2. Winner-take-all effect in a social signaling competition. Arrows indicate the quality of friends that individuals will make at equilibrium, depending on their own quality (Dessalles 2014).*

We just reviewed a few influential scenarios that have been invoked to account for the existence of language in our species. Again, the point is not to prove them definitively wrong, but rather to show their limits as they stand. The situation at this point is that no satisfactory account of the emergence of language is yet available. *The selection pressure that led to language is still missing.*

We considered the possibility mentioned by many authors that some preadaptations may have facilitated the advent of language in our lineage. Unfortunately, these preadaptations do not make the characteristics listed in Table 1 any easier to evolve. For instance, most commonly mentioned preadaptations do not explain why children, contrary to apes, systematically point to unexpected stimuli (Table 2, last column). Tomasello explains the difference with apes by invoking a specific preadaptation, namely the human ability to establish joint attention, an ability that apes would lack. The explanation has its limits. Apes and monkeys are perfectly able to take the presence of conspecifics into account when emitting alarm calls (Zuberbühler 2006). If there had been a selection pressure for evolving it, declarative pointing would have emerged in chimpanzees, at least as an automatic behavior. But it did not, probably because referential communicative acts require some level of trust (Knight 1998; Power 1998) that is absent from the competitive social world of chimpanzees (Hare & Tomasello 2004). As we can see, the very notion of preadaptation is of little help to explain the emergence of language, even in the embryonic form of declarative pointing.

The lesson of Table 2 is not only that language is left unexplained, but that it shouldn't even exist! Language as we know it should have been counter-selected. Each minus sign is Table 2 means a selection pressure in the opposite direction. Even if language had emerged as a cultural habit (Tomasello 1999a), biological evolution would have favored individuals refraining from using it the way human beings do. In other words, *language seems to be evolutionarily unstable!* Or perhaps some piece is missing in the puzzle. The next section proposes to combine the two best candidates of Table 2, namely the social networking and social signaling scenarios, to reach a convincing candidate for the missing selection pressure.

## 5. Language as a social signaling device

The hypothesis that will be further explored now is that language evolved as a form of social display. This idea comes naturally to mind once one realizes that human speech is preferentially directed toward acquaintances and that friends' most regular activity is chatter (Dunbar 1996). Individuals establish social bonds or decide to break them based on the interestingness of their social partners' conversation. The human form of communication stands out by its costs and its exaggerated amplitude (Table 1) among primate communication systems. This also points out to some kind of signaling.

After Zahavi's (1975) initial proposal, several successive models (Grafen 1990), (Gintis, Smith & Bowles 2001), (Dessalles 2014) proved the consistency and the robustness of the signaling mechanism. In all these models, individuals are represented by agents that differ by some quality. This quality remains concealed but may be revealed by sending a signal. Sending the same signal level is supposed to be more costly for low-quality individuals than for high-quality ones. In social signaling games, agents benefit from establishing bonds with higher-quality individuals. It is therefore in their interest to pay attention to quality-revealing signals, and it is in the interest of signalers to invest in those signals, even if they are costly, as it is the best way to attract allies. At equilibrium, high-quality individuals turn out to be the most intense signalers (Gintis, Smith & Bowles 2001). This means that signals become reliable indicators of qualities.

This mechanism offers new explanations for various situations of apparent altruism in animals (Zahavi & Zahavi 1997) and for the existence of prosociality in humans (Bliege Bird & Smith 2005; Hawkes 1991; Hawkes & Bliege Bird 2002; Lyle & Smith 2014; Palmer & Pomianek 2007). Though the role of language has sometimes been evoked in relation to social signals (Knight 2008; Smith 2010), the possibility that language itself could be a social signaling device has rarely been considered (if we except the case of sexual signaling, see above). Some time ago I proposed that language was used to advertise one's ability to be relevant (Dessalles 1998). If the social signaling schema is correct, then relevant speakers would get social status (Burling 1986). Unfortunately, this schema does not work because of the aforementioned winner-take-all effect (figure 2). This effect characterizes any social signaling situation in which affiliations are unconstrained. One extreme example is the Twitter social network in which some individuals may attract millions of followers while the vast majority of users have less than a dozen of active followers (Kwak, Lee & Park 2010). Large social networks are not monolithic. They are segmented into a multitude of communities sharing some common interest. The winner-take-all effect is however expected to operate in each of these communities, as long as affiliations are unconstrained: eventually, a minority of individuals will attract most of the followers by investing significantly in communication, while the majority will be unable to match the same level of signal and will be discouraged from signaling. This schema does not describe language as we know it. Virtually all people do talk (Mehl *et al.* 2007). If language is a competition for relevance, how can we explain that most individuals still strive to make verbal contributions, knowing that they cannot outcompete those who are able to entertain any audience with their brilliant conversation?

More recently I was able to design a model in which generalized social signaling emerges and remains stable (Dessalles 2014). The crucial assumption is that social bonds imply sharing time together. This time-sharing constraint tends to make relationships symmetrical. When two agents A and B meet, a negotiation takes place. A considers accepting B as friend, based not only on the signal sent by B, but also on the amount of time B offers to share (or, equivalently, based on the rank A may occupy in B's list of friends: best friend, second best friend…). B performs the same computation, and if A and B come to an agreement, they become friends. This may result in some older friend being displaced in A's or B's list of friends. The apparent quality displayed through signals is no longer the only criterion for forming alliances. Individuals base their choice on each other's 'social offer', which corresponds to signal×time (signal multiplied by the amount of time offered to share).

Under the time-sharing hypothesis, agents benefit from establishing alliances with relevant individuals, not the most relevant ones, but the most relevant among the available ones. Due to the time-sharing constraint, social competition rapidly leads to assortativity: the most relevant individuals become friends with each other, the second most relevant individuals become acquainted among themselves as well, and so on. Even the least relevant individuals have a chance to form social relationships with individuals like themselves (Figure 3). This emerging order has a consequence: each individual benefits from investing in communication to attract social partners.

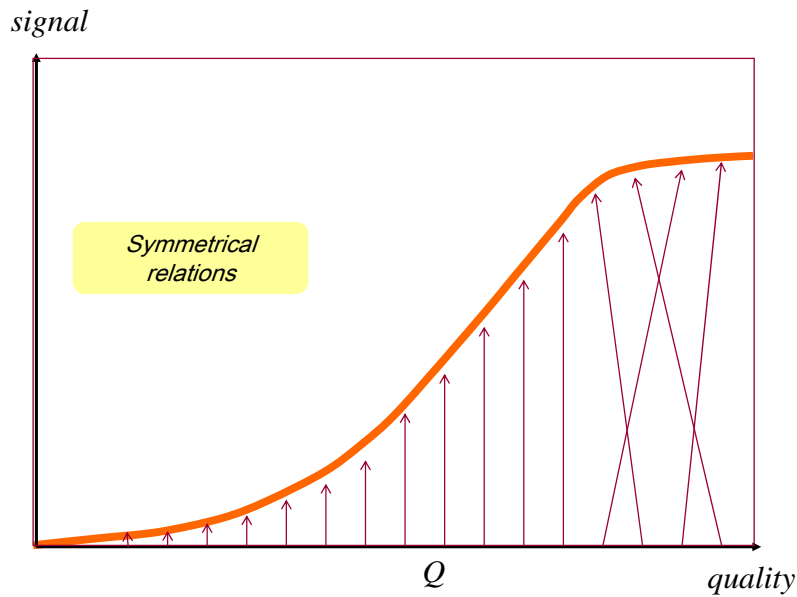

*Figure 3. Assortment in a time-sharing signaling competition. At equilibrium, individuals become acquainted with individuals of similar quality (Dessalles 2014).*

Thanks to the time-sharing assumption, there is no longer any winner-take-all effect. No one can get millions of friends as there is not enough time to give to each of them. Competition on the social market generates assortativity (Figure 3). Signals are moreover reliable indicators of quality. They end up being an increasing function of quality[8], and so the quality can be retrieved from the signal (Dessalles 2014). This provides the first schema in which a whole population of individuals finds an interest in investing in costly social communication, rather than just a minority of them.

Does this time-sharing version of the signaling model offer a plausible (though simplified) image of the role of language? The answer depends on whether its hypotheses do apply to human societies:

- Do friends share time together?
- Do we observe social assortativity?
- Is language performance involved in the selection of friends?
- Which are the social qualities advertised by language performance?
- How are theses qualities related to survival and reproduction?

## *5.1. Time-sharing*

Time sharing among friends is commonly observed among primates, in fission-fusion social organizations (Kummer 1997; Sueur et al. 2011). Populations split into smaller units on a regular basis, typically to secure access to food. Individuals actively choose with which other members of the same population they will spend the next hours or days. Hunter-gatherer societies are organized according to a similar schema. Individuals choose partners on a daily basis to forage and on a monthly basis to form camps (Marlowe 2005). In our modern societies, the sharing of activities is also crucial in friendship (Douvan & Adelson 1966). Dunbar (1996) drew attention to the analogy between language and grooming. Both activities are time consuming. By their very nature, they are reliable indicators of the fact that social partners do spend time together.

---

[8] The optimal social signal $s(q)$ that an agent with quality $q$ emits at equilibrium is given by $s'(q) = q\, P'(q)/C$, where $C$ is the cost coefficient (what it costs to increase signal intensity by one unit) and $P(q)$ is the payoff provided by affiliating with other agents with similar quality $q$. The equation relates the variation of the signal to the variation of the payoff in a quite simple way. If individuals have time for only one friend, optimal signals increase as the square of quality. For several friends, the curve is more complicated but the overall shape remains similar (Dessalles 2014).

## *5.2. Assortativity*

Social assortivity is manifest in human societies (Verbrugge 1977). Friends tend to be matched by age, education and occupational prestige. It would be interesting to investigate whether these correlations, rather than being a mere consequence of proximity, result from the search for mutual interest (or relevance) in verbal interactions (Aries & Johnson 1983). Many people hold certain individuals they do not know personally in high esteem, for instance celebrities. No doubt they would like to become acquainted with these valued figures if only they could[9]. However, due to the limitation of social time to share, individuals in great demand can only accept a few friends, and eventually friendships concern individuals of similar status.

## *5.3. The role of language in friendship*

As intuition suggests, there is a strong correlation between friendship closeness and the frequency and quality of language interactions (Dunbar 1996; Eagle, Pentland & Lazer 2009). Friends may speak about any topics, but close friends are more prone to exchange intimate and emotional experiences (Aries & Johnson 1983; Rimé 2005, p. 130). Superficially, the correlation may seem to result from a mere reinforcement: we are more likely to speak with friends as we share time with them, and we are more likely to getting acquainted with people we happen to talk to. This description at surface level is hiding the fact that a genuine social selection operates during language interactions. People choose each other based on the relevance (or interest) of their conversation. We seek the company of people who are able to stir up our interest and conversely we tend to turn away from individuals whose conversation is repeatedly boring or trivial. Failures in connecting topics or ideas are regarded as pathological (Meilijson, Kasher & Elizur 2004) and expose individuals to social exclusion.

## *5.4. Qualities advertised by language performance*

The signaling model relies on the assumption that language is used to display qualities. What are they? I proposed that the purpose of human conversation is to advertise one's ability to be relevant (Dessalles 1998). In our species, relevance and interest are primarily linked to unexpectedness and abnormality (section 3). What is at stake in spontaneous conversations is not the topic itself, which is most of the time inconsequential to the participants (Dunbar, Duncan & Marriott 1997). Many situations, even the most inconsequential ones (e.g. a third-division team reaching the semi-finals) can be used as an excuse to signal something unexpected. It seems that the proximal purpose of language is to point to abnormal situations and discuss about them, whatever the subject matter. During story rounds (Tannen 1984, p. 100) individuals literally compete for recounting unexpected events, and during discussions, participants compare their ability to deal with problematic issues (Dessalles 2016; 2017).

---

[9] In previous analyses (Dessalles 1998), I considered a reified version of social status and tried to make it the outcome of language interactions: individuals would grant a bit of social status to relevant people. This schema proved to be a non-starter. The present model focuses on the relation between being relevant and chances of establishing social bonds. Social status becomes nothing more than an emergent property: the status of some individuals emerges from the wish of many to become acquainted with them.

The dimension of abnormality seems to be systematically present in spontaneous conversation topics. However, another dimension may amplify interest and relevance: the ability to share emotion about the situation which is recounted or discussed (Rimé 2005, p. 114). Not only do people use language mostly to elicit surprise and emotion, but their audience enjoys being surprised and emotionally moved (Rimé 2005, p. 177).

What speakers demonstrate during verbal interaction is the ability to deal with abnormality, either by pointing to it or by discussing about it. Since information can be defined in terms of unexpectedness[10], we can say that individuals advertise their *informational* abilities. For the signaling model to apply to language and depict it as a social display mechanism, we must now understand how this abnormality-oriented signaling activity is related to survival and reproduction, both on the speaker's side and on the listener's side.

## 6. The missing selection pressure: one candidate

Table 2 suggests that none of the common evolutionary scenarios can explain, not only the emergence, but also the persistence of language as it exists in our species. In the previous section, we considered the possibility that language could be an instance of social signal. Individuals would advertise their informational capacities, *i.e.* their ability to deal with abnormal situations by chatting about them. This proposal ticks all the boxes: it explains why individuals have an incentive to talk (SPADV) competitively (OFFER) to anyone (NOAUDEX), even if it requires time and efforts (COST), since it is a way for them to build their social network; it explains why people talk quite often about inconsequential matters (INCONS), since the point is to deal with abnormality, taking every opportunity to do so; tanks to the time-sharing hypothesis, all individuals have an incentive to enter the signaling competition (GEN, GDRN); the use of plethoric vocabularies (VOCAB) is in part explained by the fact that unexpected situations are by definition rare and need precise descriptions to be distinguished (Briscoe 2006).

One crucial feature remains to be explained: why does human conversation systematically deal with abnormal states of affairs (ABNORM)? Why do human beings find an interest in establishing social bonds with individuals who are best able to point to unexpected, abnormal situations and to discuss about them? This question constitutes the keystone of the signaling model of language. Finding an adequate answer proved harder than anticipated, until the hypothesis that we will consider now emerged as a natural answer.

---

[10] Shannon's definition of information refers to surprise, which is quantified using the improbability of events. This definition has been extended within the Algorithmic Information Theory framework and is quantified as abnormal simplicity (see Dessalles 2013).

### 6.1. Why are abnormal situations relevant?

To explain why human conversations are systematically abnormality-oriented (section 3), I propose to start, not from a hypothesis, but from a fact. At some (unknown) point in the hominin history, individuals started using weapons to commit risk-free homicide. This mere fact must have had dramatic consequences on earlier social organizations.

*In normal circumstances the possession by all men, however physically weak, cowardly, unskilled or socially inept, of the means to kill secretly anyone perceived as a threat to their own well-being not only limits predation and exploitation; it also acts directly as a powerful levelling mechanism. (Woodburn 1982)*

We are the only social lineage in which risk-free killing became possible. Any previous social organization based on coercion collapsed with such an option offered to all members. The event generated a political singularity. It did not only provoke a transition to egalitarian societies (Boehm 1999; 2008), but also a complete change of survival strategy. Almost overnight, hominin individuals represented a mortal danger to their group mates. Human violence, as observed in various culture including hunter-gatherers, is characterized by less aggression (Wrangham, Wilson & Muller 2006) but a similar death rate in comparison with apes (Gómez, Verdú & González-Megías 2016), due to massive intra-group interpersonal homicide with only one perpetrator (Fry & Söderberg 2013; Hill, Hurtado & Walker 2007). The date of the weapon singularity is unknown, but it is suspected to have occurred early in the hominin ancestry (Sala, Arsuaga & Pantoja-Pérez, 2015).

The weapon singularity has rarely been acknowledged as a crucial event in hominin history (see however Bingham 2001). Far from being a banal episode on the route to *Homo sapiens*, it might well be a bifurcation point that had several decisive consequences. One is the social leveling mentioned by Woodburn and Boehm. Another one is that individuals become eager to discover hidden facts about each other (Locke 2005; 2010). But the most important logical consequence of the weapon singularity for our concern here is that criteria for social bonding must change. When danger is unpredictable and may come from group mates, from friends and even from within the couple, choosing the right social partners becomes a question of survival. Ideal friends should help you anticipate danger, they should be ready to spend time with you and they should not be themselves a danger to you.

After the weapon singularity, information replaced muscle: informed friends became more valuable than strong ones. Language as we use it could be a remote consequence of this shift in social values. By taking any opportunity, even the most futile ones, to signal and discuss abnormal situations, people display their ability to spot potential danger. People we find interesting are those who are able to surprise us by recounting unexpected events. Individuals who are best able to demonstrate their vigilance by doing so attract more social interest than individuals who seem to know nothing noticeable about the surrounding physical or social world.

This possible link between the weapon singularity and the emergence of language provides a consistent scenario that passes all the tests considered above (Table 2). Individuals have a vital interest in sharing time with alert friends who are the most capable of anticipating homicide risk. They make every attempt to attract these friends by displaying their own ability to spot the unexpected. The missing slot in the last row of Table 2 (GEN) is now filled: since protection is only efficient as friends spend time together, the time-sharing model applies and all individuals have an incentive to display their informational abilities.

The weapon singularity scenario not only explains why our conversations are oriented towards abnormality (section 3), but also why we tend to share emotional events with close friends (Aries & Johnson 1983; Rimé 2005, p. 130) and why, conversely, we tend to avoid being closely acquainted with people who show no emotions or show emotions opposite to ours (Rimé 2005). This fundamental aspect of human verbal interactions makes sense within the weapon singularity scenario: by revealing the conditions in which they feel emotions, individuals become more predictable and appear less likely to represent a danger themselves.

## 6.2. *From pointing to syntactic language*

The evolutionary scenario outlined in the previous section explains how a new selection pressure may have eventually led to language. The transition to language is unlikely to have been instantaneous, though. Our languages rely on complex recursive structures and on the use of predicates (section 3). Within utilitarian scenarios of language evolution, complex features like this seem absurd (Premack 1985, p. 282; Redhead & Dunbar 2013). What is the selection pressure that led to such sophistications, and how are they adapted to the corresponding function? It is beyond the scope of this article to discuss these issues in detail. My hypothesis is that the main selection pressure remained constantly the same: establish protective social bonds by advertising one's ability to anticipate unexpected violence from other group members. I proposed that the capacity to form predicates and to reason logically emerged in a further step, initially as a protection against lies (Dessalles 1998). The point is not to insure epistemic quality, as has been suggested (Sperber 2000; Mercier & Sperber 2011), since most discussions are about inconsequential topics. People raise consistency issues publicly through argumentation[11], not to improve the quality of their own knowledge, but to advertise their ability to distinguish genuine abnormality[12] (Dessalles 2011a). Once the cognitive ability to form predicates was there, syntax then evolved as an efficient means to express them (Dessalles 2007b[2000], pp. 216-20).

This paper had two objectives. The first one was to show that determining the selection pressure that led to language is a genuine scientific problem which is by no means "obvious" and which still remains unsolved. In particular, scenarios relying on collective advantage, on coordination or on reciprocal cooperation are unable to provide satisfactory answers to the problem. The second objective was to propose a logically consistent solution, in which language appears as a social signaling device. People would advertise their social value by talking about abnormal situations. Thanks to the time-sharing requirement which comes naturally with the protective scenario, we avoid the winner-take-all effect that characterizes other social signaling systems. All individuals have an incentive to advertise their ability to spot and to discuss the unexpected, as a way to demonstrate their skill at anticipating dangerous situations. Human language would be a far consequence of an event that did occur in the past of our lineage and that I named the weapon singularity. This event provides a clear-cut reason why language is unique to our species.

---

[11] See (Dessalles 2016) for a cognitive model of argumentative competence.

[12] A side-effect of this propensity to denounce inconsistencies is that people can talk about distant events with marginally costless signals (Scott-Phillips 2007). I suggested that before the emergence of this behavior and of the corresponding abilities, only verifiable (almost-here-almost-now) events were worth signaling, and that protolanguage was a locally optimal means to do so (Dessalles 2007b[2000], p. 332).

Several aspects of the scenario can be tested. We can investigate whether human conversation is universally oriented toward abnormal states of affairs. I could verify for instance that Japanese spontaneous conversation does not significantly differ from French conversation in this respect (Dessalles 2011b). We can design experiments to show that unexpectedness systematically raises interest in adult and in children, independently from consequentiality. We can design experiments to contrast what constitutes a surprise in apes and in humans (for instance, are apes surprised by coincidences?) We can get some insight into the weapon singularity by studying historical or anthropological records about the reaction of societies when new weapons such as poison permitting anonymous killing were introduced, or by observing the way people adapt their social network when the protection of law disappears. We can also investigate whether various cognitive components underlying language competence (quantification/determination, aspect, tense, contrast and predication (Dessalles 2015), abduction, negation (Dessalles 2016)) can be shown to be well-adapted to serve the purpose of signaling and discussing abnormal situations (for instance, aspectual relations can be crucial to support negation, as in the constitution of an alibi, and thus to point to an inconsistency).

One of the most characteristic behaviors of our species, language, still remains an evolutionary mystery. Actively and repeatedly drawing other individuals' attention to abnormal states of affairs seems pointless from a Darwinian point of view. What benefit do speakers get from doing so? In an attempt to solve this conundrum, I suggested that language is not a tool for exchanging useful information, but rather a way of displaying one's ability to acquire information. I showed how human conversation behavior can be an evolutionary stable strategy as soon as informational abilities are regarded as a social signal, *i.e.* a signal used to attract friends, in a context in which friends share time together. To explain why informational abilities play such a central role in hominin societies, I proposed to establish a link with the weapon singularity. Far from being an anecdotal episode in hominin history, the weapon singularity may be a determining point from which our lineage diverged from other primate social organizations. In this new social niche, individuals are seeking informed friends and spend hours showing off their own informational abilities. Language as we know it may have emerged to fulfill that function.

## Acknowledgments

I am indebted to Giovanni Sileno for his incisive and fruitful remarks on earlier versions of this paper. I would like to thank Chris Knight and Jim Hurford for having encouraged me to develop this line of research and for the confidence they put in me from the start.